\documentclass{elsart}
\usepackage{graphicx}

\begin{document}

\def\dj{\leavevmode\setbox0=\hbox{d}\dimen0=\wd0
        \setbox0=\hbox{d}\advance\dimen0 by -\wd0
        \rlap{d}\kern\dimen0\hbox to \wd0{\hss\accent'26}}
\def\DJ{\leavevmode\setbox0=\hbox{D}\dimen0=\wd0
        \setbox0=\hbox{D}\advance\dimen0 by -\wd0
        \rlap{D}\kern\dimen0\hbox to \wd0{\raise -0.4ex\hbox{\accent'26}\hss}}

\begin{frontmatter}

\title{Cluster structure of $^{13}$C probed via the $^7$Li($^{9}$Be,$^{13}$C$^*\rightarrow^{9}$Be+$\alpha$) reaction}

\author[bham]{N. Soi\'{c}\thanksref{irb}},
\author[bham]{ M.~Freer},
\author[bham]{ L. Donadille\thanksref{sac}},
\author[bham]{ N. M. Clarke},
\author[bham]{ P. J. Leask},
\author[surrey]{W. N. Catford},
\author[surrey]{ K. L. Jones\thanksref{gsi}},
\author[surrey]{ D. Mahboub},
\author[york]{ B. R. Fulton},
\author[york]{ B. J. Greenhalgh},
\author[york]{ D. L. Watson} \and
\author[anu]{D. C. Weisser}

\address[bham]{School of Physics and Astronomy, University of Birmingham, Edgbaston,
Birmingham B15 2TT, U. K. }
\address[surrey]{School of Electronics and Physical Sciences, University of Surrey, Guildford,
 Surrey, GU2 5XH, U. K.}
\address[york]{Department of Physics, University of York, Heslington, York, YO10
5DD, U. K.}
\address[anu]{Department of Nuclear Physics,  The Australian National University, Canberra ACT
 0200, Australia }

\thanks[irb]{Permanent address: Ru\dj er Bo\v{s}kovi\'{c} Institute, Bijeni\v{c}ka 54, 
HR-10000 Zagreb, Croatia}
\thanks[sac]{Present address: CEA-Saclay, DAPNIA/SPhN, Bt. 703, Pice 162, F-91191 Gif sur
 Yvette Cedex, France}
\thanks[gsi]{Present address: GSI, Gesellschaft f\"{u}r Schwerionenforschung mbH,
 Planckstrasse 1, D-64291 Darmstadt, Germany}

\begin{abstract}
A study of the $^7$Li($^{9}$Be,$^4$He$^{9}$Be)$^{3}$H reaction at
$E_{beam}$=70 MeV has been performed using resonant particle
spectroscopy techniques and provides a measurement of
$\alpha$-decaying states in $^{13}$C. Excited states are observed
at 12.0, 13.4, 14.1, 14.6, 15.2, 16.8, 17.9, 18.7, 21.3 and 23.9
MeV. This study provides the first measurement of the three
highest energy states. Angular distribution measurements have been
performed and have been employed to indicate the transferred
angular momentum for the populated states. These data are
compared with recent speculations of the presence of chain-like
structures in $^{13}$C.
\end{abstract}


\end{frontmatter}

\section{Introduction}
The structure of light nuclei has always provided a certain
fascination, given that a full range of structural properties are
displayed from spherical shell model like structure via
deformation to clustering. In particular even-even nuclei,
composed of equal numbers of protons and neutrons offer the
possibility that the nucleus may be decomposed into a collection
of $\alpha$-particles. This was the picture developed by Ikeda
\cite{Ike68}, who speculated that at the point that the $N\alpha$
decay threshold was encountered the associated cluster degree of
freedom should be liberated. Correspondingly, this picture would
suggest that the $^8$Be ground state should possess an
$\alpha$-$\alpha$ cluster structure, and the 3$\alpha$ cluster
structure should appear at an excitation energy close to 7.3 MeV
in $^{12}$C. Indeed, it has been recently speculated that certain
states above the 3$\alpha$ decay threshold in $^{12}$C (e.g. the
7.65 MeV state), or the 4$\alpha$ decay threshold in $^{16}$O,
should possess properties reminiscent of a Bose gas \cite{Toh01}.

These highly developed cluster systems have been shown to have an important
impact on the structure of neutron-rich isotopes of beryllium and carbon. For
example, in the case of $^9$Be the properties of the valence neutron may be
explained in terms of the sharing of the neutron between the two $\alpha$-cores
in a manner which is reminiscent of the covalent binding of atomic molecules
\cite{von96,von97,von97a}. For valence neutrons based on $\alpha$-particle
cores, linear combinations of the $p$-type orbits lead to analogues of the
atomic $\sigma$ and $\pi$-bonds. Indeed, such a description provides a useful
basis for the understanding of the low lying spectroscopy of the $^{9}$Be
nucleus, with the ground state structure forming a rotational band with
$\pi$-like character.

The presence of 3$\alpha$ cluster structures in $^{12}$C provides
an extension to the ideas developed for the two-centre systems.
Calculations which explicitly contain the molecular-orbital
structure \cite{Ita01} show that unlike the linear 3$\alpha$
system, which is unstable against bending modes and thus collapse
into more compact structures, the 3$\alpha+Xn$ systems are stable
against the bending mode \cite{Ita01,Ita02}. Thus, there would
appear to be an opportunity for observing linear chain
configurations, partially stabilized by the valence neutrons in
covalent type orbits. A recent analysis of available data by Milin
and von Oertzen has provided tentative evidence for the existence
of such a structure in $^{13}$C \cite{Mil02}. Via a careful
examination of previous experimental spin and parity assignments,
two rotational bands were proposed ($K=3/2^+, 3/2^-$) whose
deformations indicated a chain-like structure and which were split
in energy due to the intrinsic asymmetry of the 3$\alpha+n$
structure. This analysis provided information on such rotational
structures up to $E_{x}\sim$17 MeV. Although, such a picture is
appealing the relationship between the states considered by Milin
and von Oertzen has not been demonstrated since the spins and
parities remain to be confirmed.

In the present paper we present a measurement of the
$^{7}$Li($^9$Be,$^{13}$C$^*\rightarrow^{9}$Be+$\alpha$)$^{3}$H
reaction. The $\alpha$-transfer process provides a possible mechanism
by which the chain-like structures in $^{13}$C may be populated. The
measurement covers the excitation energy range of 12 to 28 MeV,
the interval over which the higher energy and spin members of the
two rotational bands are believed to exist.

\section{Experimental Details}

The measurements were performed at the Australian National University's 14UD
tandem accelerator facility. A 70 MeV $^9$Be beam, of intensity 3 enA, was
incident on a 100 $\mu$gcm$^{-2}$ Li$_2$O$_3$ foil. The integrated beam
exposure was 0.45 mC.

Reaction products formed in reactions of the $^9$Be beam with the
target were detected in an array of four charged particle
telescopes. These telescopes contained three elements which
allowed the detection of a wide range of particle types, from
protons to $Z$=4 to 5 nuclei. The first elements were thin,
70$\mu$m, 5$\times$5 cm$^{2}$ silicon detectors segmented into
four smaller squares (quadrants). The second elements were
position-sensitive strip detectors with the same active area as
the quadrant detectors, but divided into 16 position-sensitive
strips. These strips were arranged so that the position axis gave
maximum resolution in the measurement of scattering angles.
Finally, 2.5 cm thick CsI detectors were used to stop highly
penetrating light particles. These detector telescopes provided
charge and mass resolution up to Be, allowing the final states of
interest to be unambiguously identified. The position and energy
resolution of the telescopes was $\sim$1 mm and 300 keV,
respectively. Calibration of the detectors was performed using
elastic scattering measurements of $^{9}$Be from $^{197}$Au and
$^{12}$C targets. The four telescopes were arranged in a
cross-like arrangement, separated by azimuthal angles of 90$^{\circ}$.
Two opposing detectors were located with their centres at
17.3$^{\circ}$ and 17.8$^{\circ}$ (detectors 1 and 2) from the 
beam axis and with the strip detector 130 mm from the target, whilst the
remaining pair were at the slightly larger angles of 28.6$^{\circ}$
and 29.7$^{\circ}$ (detectors 3 and 4), 136 mm from the target.

\section{Results}

\begin{figure}
\includegraphics[width=0.75\textwidth]{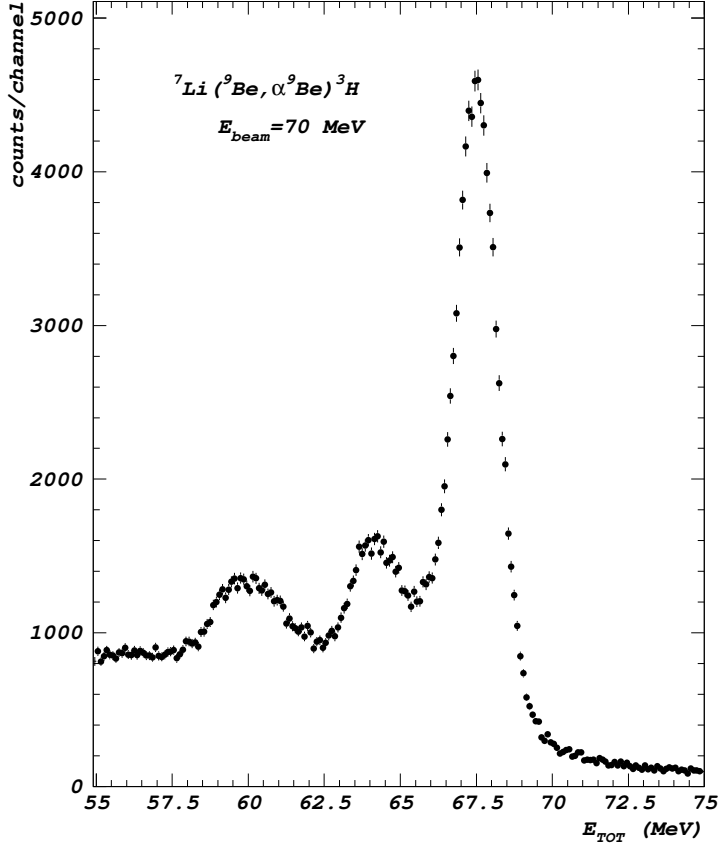}
\caption{\label{etot}
Total energy spectrum for $^{4}$He+$^{9}$Be coincidences detected
in telescopes 1 and 2, assuming a
$^3$H recoil. The highest energy peak corresponds to the observation
of the $^7$Li($^9$Be,$^9$Be$\alpha$)$^3$H reaction. Statistical
uncertainties are presented only.}
\end{figure}

The detection system employed in the present measurements allowed
a measurement of the charge, mass, energy and emission angle (and
hence a determination of the momentum) of each particle. This
provides a complete determination of the reaction kinematics in
the case of a three particle final state. Figure \ref{etot}
shows the total energy spectrum of the three final state particles
assuming a $^3$H recoil ($E_{tot}=E_{Be}+E_{\alpha}+E_t$),
after gating on $^9$Be+$\alpha$ in the particle identification
spectra. The large
peak in this spectrum corresponds to the observation of the
$^{7}$Li($^{9}$Be,$^9$Be$\alpha$)$^{3}$H reaction (Q=-2.468 MeV),
where all particles are produced in their ground states ($^9$Be
only possesses one state bound against particle decay). The energy
resolution in the total energy spectrum is $\sim$1.9 MeV,
considerably worse than the intrinsic energy resolution of the
detectors (300 keV). This is due to in fact the uncertainty
imposed by the measured momenta of the two detected particles
limits the resolution of the reconstructed momentum of the
undetected triton. Given the relatively small mass of the recoil
particle the uncertainty in momentum is magnified in energy. The
first peak at lower total energy corresponds to reactions
involving the oxygen and carbon components in the target:
$^{16}$O($^9$Be,$^{9}$Be$\alpha$)$^{12}$C (Q=-7.162 MeV) and
$^{12}$C($^9$Be,$^{9}$Be$\alpha$)$^{8}$Be (Q=-7.367 MeV),
predominantly the former, with the lowest energy peak coinciding
with the 4.4 MeV excitation of the $^{12}$C recoil.
Unfortunately, the underlying background means that these
channels cannot be analyzed.

\begin{figure}
\includegraphics[width=0.75\textwidth]{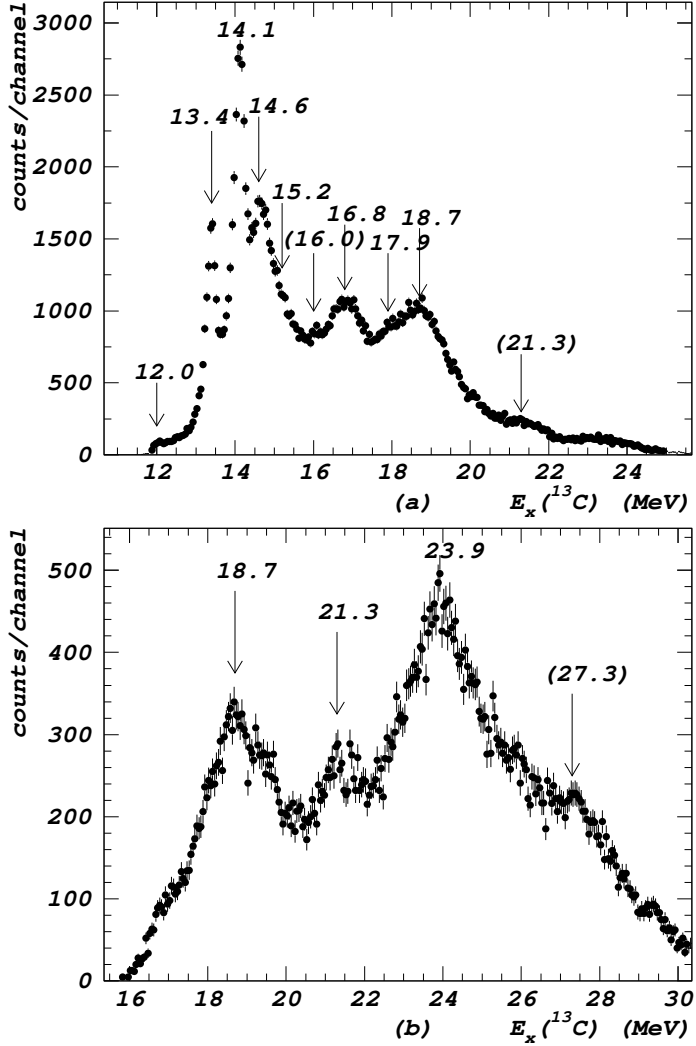}
\caption{\label{exc}
$^{13}$C excitation energy spectra for decays detected in 
detectors 1 and 2 (a) top, and 3 and 4 (b) bottom. Observed peaks
are labelled with excitation energies (MeV). The details of the
peaks observed in these spectra are presented in Table \ref{13cexc}.
Error bars represent statistical errors.}
\end{figure}

By selecting only the events associated with the
$^{7}$Li($^{9}$Be,$^9$Be$\alpha$)$^{3}$H reaction the $^{13}$C 
excitation energy can be calculated from the relative velocity of 
the two decay fragments. Figures \ref{exc}a (top) and \ref{exc}b 
(bottom) show such spectra for the small angle detector
pair (1 and 2) and the larger angle pair (3 and 4) respectively. 
As expected these spectra sample slightly different excitation energy 
ranges due to the minimum opening angle required for the fragments to 
reach the detectors. Given the presence of three particles in the final 
state the possibility remains that at least some of the yield in 
Figures \ref{exc}a and \ref{exc}b arises from decays of either $^7$Li 
into t+$\alpha$ or $^{12}$B to t+$^{9}$Be. Both of these
decay processes were reconstructed and only evidence for decays 
of $^7$Li was found, where the states at 4.63 and 6.68 MeV were 
observed in the coincidence spectra for detectors 3 and 4. This 
contribution was removed from the spectrum in Figure
\ref{exc}b by eliminating all data corresponding to $^7$Li 
excitations less than 7 MeV.

\begin{table}
\caption{\label{13cexc}
 Excitation energies of $^{13}$C states $\alpha$-decaying into the
ground state of $^{9}$Be. The uncertainty in these energies is 100 keV. The
previous measurements are from the tabulations of \cite{Ajz91}. Note that the
excitation energy resolution is 180 to 200 keV from an analysis of the 13.4 MeV
peak, and therefore widths are only quoted where the experimental resolution
permits, and the experimental resolution has been unfolded. The states shown in
italics are states observed at the same energies as those in the present
measurements, but have different experimental widths.}

\begin{tabular}{|c|c|c|c|c|}\hline
\multicolumn{2}{|c|}{Present} & \multicolumn{3}{|c|}{Reference \cite{Ajz91}}\\

$E_x$  & Width  & $E_x$  & $J$ & Width
\\
(MeV) &(keV) &(MeV) & & (keV) \\
\hline

12.0 & & 11.95 & 5/2$^+$ & 500 \\

13.4 & & 13.41 & (9/2)$^-$ & 35\\

14.1 & & 14.13 & 3/2$^-$ & $\sim$150\\

14.6 & & 14.58 & (7/2$^+$, 9/2$^+$) & 230\\

(15.2) & & 15.27 & 9/2$^+$ & \\

(16.0) & & 16.08 &(7/2$^+$) & 150\\

16.8 & 310 & 16.95  & & 330\\

(17.9) &    & 17.92 & & \\

18.7 & 570 & {\em 18.70} & {\em (3/2$^+$, 5/2$^+$)} & {\em 100}
\\
21.3 & 530  & {\em 21.28}  & & {\em 159} \\

23.9 & 1100 & {\em 24.00} & & {\em 4000} \\

(27.3) & & {\em 27.5} & &
\\ \hline
\end{tabular}
\end{table}

 The resulting excitation energy spectra extend from the threshold 
at 10.65 MeV to $\sim$28 MeV. Characteristically, the states are 
narrow at low energies, becoming broad at the other limit. The 
excitation energies and widths of these states are presented in 
Table \ref{13cexc}. In the region of overlap between the
two spectra it is clear that there is good agreement with in 
particular the 18.7 MeV peak being represented in both spectra, 
and to a lesser extent the 21.3 MeV peak. The diminished strength 
in the spectrum in Figure \ref{exc}a at high excitation energies is 
related to the decrease in the detection efficiency. The spectra show
a number of strong peaks at  13.4, 14.1, 14.6, 16.8, 18.7, 21.3 
and 23.9 MeV, and there is some evidence for additional states at 12.0, 
15.2, 16.0, 17.9 and 27.3 MeV. These weak states are not clearly 
observed in the spectra in Fig \ref{exc}, which represent the total sum
of good events recorded by telescopes 1-2 and 3-4, but were observed in
some individual spectra for the coincidences and are moreover required
to provide a good reproduction of the shape of the spectra with multiple 
peak and a linear background fitting. The experimental excitation energy 
resolution can be observed from the width of the 13.4 MeV peak to be 
180 to 200 keV. This limit only permits widths of the broader states 
to be quoted. This has been done in Table \ref{13cexc} by deconvolving the
experimental resolution assuming that the experimental and physical width 
add in quadrature. The energy dependence of the excitation energy resolution 
has been calculated and suggests moderate change of the excitation energy
resolution with energy: at 17 MeV excitation the resolution is 230 keV, 
at 19 MeV it is 240 keV and at 24 MeV it is 340 keV. Given the difference
in the widths of the resonances observed at 18.7, 21.3 and 23.9 MeV and 
those appearing in the tabulations \cite{Ajz91}, it is probable that these 
peaks correspond to new resonances in this nucleus, which may possess a 
structural (rotational) link with those at lower energies.

\begin{figure}
\includegraphics[width=1.\textwidth]{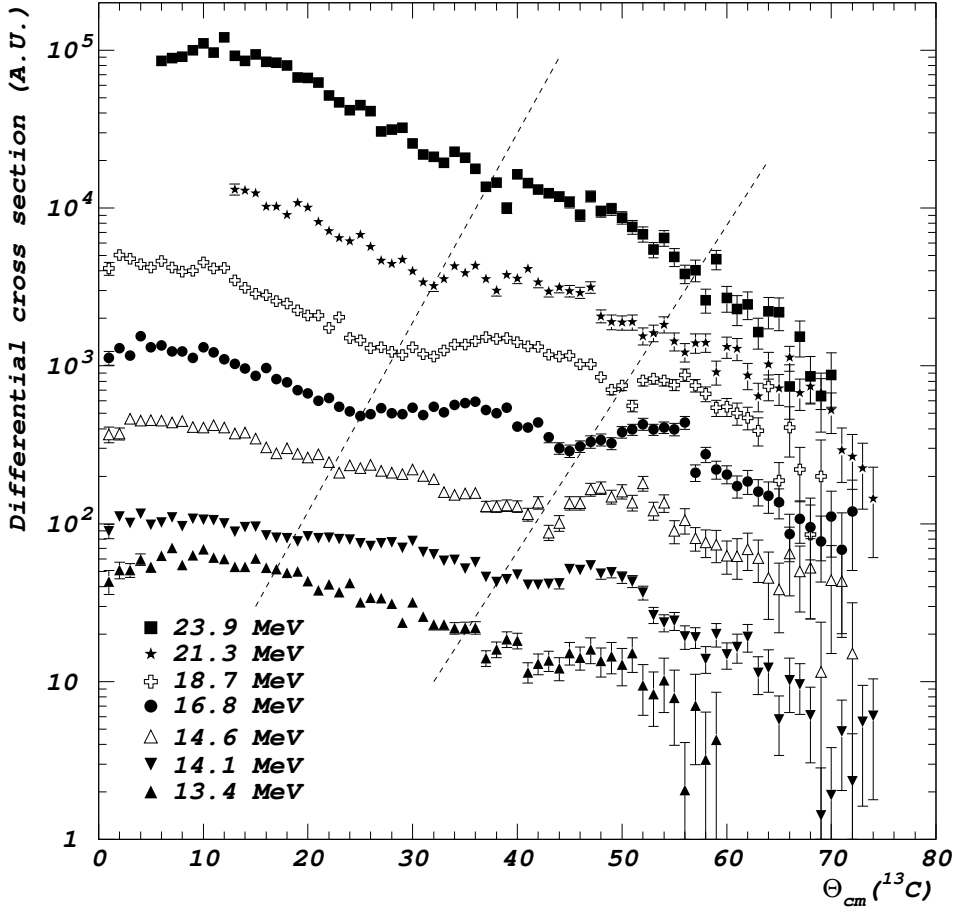}
\caption{\label{expad}
The experimental angular distributions ($d\sigma/d\Omega$) for
the states observed between 13 and 24 MeV. The distributions have been
normalised so that the data for the 7 states can be displayed simultaneously.
The dash lines indicate the location of the minima in the angular
distributions. It should be noted that for the 21.3 and 23.9 MeV states 
distributions do not extend to zero degrees due to the limit acceptance of 
the detection system in this region.}
\end{figure}

 The angular distributions of the $^{13}$C$^*$ nuclei produced in the
 $^{7}$Li($^9$Be,$^{13}$C$^*$)$^3$H reaction prior to break-up may be
 reconstructed from the measured momenta of the detected $^9$Be+$\alpha$
 decay products. These angular distributions ($d\sigma/d\Omega$)
 are shown plotted as a function of centre-of-mass emission angle of the
 $^{13}$C nucleus in Figure \ref{expad} for the main states observed in the
 excitation energy spectra in Figure \ref{exc}. It should be noted that the
 angular distributions have been normalised arbitrarily so that all the
 distributions may be plotted in the same figure.
 The distributions have been corrected for the efficiency of the
 detection of the $^{13}$C$^*\rightarrow^9$Be+$^4$He decay process
 using Monte Carlo simulations. Due to the decay of the
 $^{13}$C nucleus into two fragments which then are detected at large angles,
 it is possible to measure the angular distributions all the way down to
 zero degrees. At the other limit these distributions extend up to
 centre-of-mass angles of $\sim$80$^{\circ}$. These distributions are typical of
 such heavy-ion reactions in that they are not highly structured. Nevertheless,
 for certain of the distributions, most notably those associated with the 14.1
 to 21.3 MeV peaks there is some oscillatory behaviour. The curious feature of
 the structured angular distributions is that the minima shift with increasing
 excitation energy in a very systematic fashion, as indicated by the dash lines.
 An examination of the angular correlations, using the techniques outlined in
 \cite{Fre96}, shows that there is no structure in the distributions in the
 decay reference frame. This is as expected for non-zero spin particles
 involved in the reaction, and is due to the presence of the large number of
 reaction amplitudes contributing to the excitation and decay processes.

 \section{Discussion}

 It is useful to draw a comparison with the present measurement 
and that also using the $^{7}$Li($^9$Be,$^{9}$Be$\alpha$)$^{3}$H 
reaction at $E_{beam}$=58.5 MeV \cite{Lee98}. This earlier study 
was much more limited in statistics and in terms of the excitation 
energy range in $^{13}$C, but found evidence for the $\alpha$-decay 
of a series of states between 10.5 and 14.5 MeV. Close to 12.0 MeV,
the 11.848 MeV (7/2$^+$) and 11.95 MeV (5/2$^+$) states were 
observed to play an important role and close to 13.4 MeV, states 
at 13.28 MeV (with a tentative assignment of 3/2$^-$) and 13.41 MeV 
(tentatively assigned 9/2$^-$). Similar resonance states were observed 
in the measurements of the $^9$Be($\alpha$,$\alpha$)$^9$Be reaction 
\cite{Gos73}. In this latter study resonant structures were observed 
corresponding to $^{13}$C excitations of 13.28, 13.42, 14.11 and 
14.63 MeV which coincide with the main features in the 
present data. Thus, there is good agreement with the present 
excitation energy spectrum and those observed in other reactions 
which probe an overlap with $^9$Be+$\alpha$.

On the other hand, the angular momentum and parity of many of
these states are far from clear, especially above 12 MeV. As an
example, the 14.13 MeV state which is recorded in \cite{Ajz91} as
possessing a firm spin and parity assignment of 3/2$^-$, from
measurements of n+$^{12}$C \cite{Tor85} scattering and p+$^{13}$C
inelastic scattering \cite{Col88}, is found in the measurements of
$^9$Be($\alpha$,$\alpha$)$^9$Be to be 5/2$^-$ \cite{Gos73} and was
recently proposed by Milin and von Oertzen \cite{Mil02} to possess
spin and parity 9/2$^-$. In this latest analysis four new
assignments have been suggested for states in $^{13}$C. The spin
and parities suggested by these latter authors are given in Table
\ref{matkovon}. The two rotational bands identified in this table
are believed to be linked with a 3$\alpha$+n chain structure in $^{13}$C.
Corresponding to the predicted positive parity band shown in this
table, the 11.950, 13.41, 15.28 and 16.95 MeV states are observed
in the present measurements, and for the negative parity band the
14.13 and 16.08 MeV states are represented. However, this
compilation does not include the 14.6 MeV state observed in both
the present measurements and those of the
$^9$Be($\alpha$,$\alpha$)$^9$Be reaction, which may be an
important omission given the strength with which the state is
observed.

 \begin{table}
 \caption{\label{matkovon}
 Excitation energies and assigned spins from the analysis of
states populated in reactions involving an overlap with an $\alpha$+$^9$Be
structure from \cite{Mil02}. The spins proposed by Milin and von Oertzen are
given in the first column and those appearing in data tabulations \cite{Ajz91}
are given in the last column. }

\begin{tabular}{ccc}\hline
$J^\pi$ & $E_x$ & $J^\pi$ from \cite{Ajz91} \\

proposed & (MeV) & \\ \hline

3/2$^-$   &  9.897 & 3/2$^-$ \\

5/2$^-$   &  10.818 & 5/2$^-$ \\

7/2$^-$   &  12.438 & 7/2$^-$ \\

9/2$^-$   &  14.13  & 3/2$^-$ \\

11/2$^-$   &  16.08 & 7/2$^+$ \\ \hline

3/2$^+$   &  11.080 & 1/2$^-$ \\

5/2$^+$   &  11.950 & 5/2$^+$ \\

7/2$^+$   &  13.41  & (9/2$^-$) \\

9/2$^+$   &  15.28 & 9/2$^+$ \\

11/2$^+$   &  16.950 & None \\ \hline

\end{tabular}
\end{table}

The angular distributions in Figure \ref{expad} appear to show two features,
the first is that the distributions for the five states at 14.1, 14.6, 16.8,
18.7 and 21.3 MeV all have a similar character, and secondly it is distinct from
that of the distributions for the 13.4 and 23.9 MeV states. This difference is in
terms of the reduced oscillatory structure for the distributions of the latter two 
states, and that the distributions for these two states appear to possess a gradient 
which differs to that of the remaining states. This feature would suggest that the
two sets of states possess different characters, either in terms of their parity or 
the angular momentum transfer. In the present case positive parity states would be 
populated by odd angular momentum $L_f$ values ($L_f$ being the angular momentum 
of the orbit into which the $\alpha$-particle is transferred), whilst negative 
parity states would be populated by even $L_f$ values. 
The spins which could be populated in the transfer of an $\alpha$-particle 
with this angular momentum would be in the range $|L_f-3/2|\le J\le L_f+3/2$. Thus, 
for example, for the sequence of states 3/2$^+$, 5/2$^+$, 7/2$^+$, 9/2$^+$ and 11/2$^+$ 
the $L_f$ values would be (1,3), (1,3), (3,5), (3,5) and (5,7) respectively.
Here $L_i$, the initial angular momentum of the $\alpha$-particle in $^7$Li, is equal 
to 1. The transferred angular momentum $L$ is given by $\mathbf{L=L_f-L_i}$, and is 
even for positive parity states and odd for those with negative parity. 

The ($^7$Li,t) reaction has a long history in the population of $\alpha$-cluster 
states in light nuclei, and the angular distributions for differing $L_f$-values 
have been collated (in particular for $L_f$ = 0 and 2) by Bethge \cite{Bet70}, 
together with those for the ($^6$Li,d) $\alpha$-transfer reaction. Typically, 
the ($^7$Li,t) angular distributions are less oscillatory than those for the 
($^6$Li,d) reaction, and are reminiscent of those that appear in Figure  \ref{expad}. 
The reduced structure in the ($^7$Li,t) angular distributions is related to the
presence of two transfer angular momentum values, $L$, for each $L_f$
\cite{Bet70,Puh70}, for $L_f>$0. 
Nevertheless, the compilation of the $\alpha$-transfer reaction in \cite{Bet70} 
indicates the sensitivity to the orbital angular momentum of the orbit into which 
the $\alpha$-particle is transferred. These data indicate that the second maximum 
for $L_f$=0 and $L_f$=2 transfers appear at centre-of-mass angles 20$^\circ$-30$^\circ$ 
and 40$^\circ$-50$^\circ$, respectively. Via the properties of Bessel functions 
the corresponding locations of the maxima for $L_f$=1 and $L_f$=3 transfers 
would appear at 30$^\circ$-40$^\circ$ and 47$^\circ$-57$^\circ$, respectively.
For the five states with similar angular distributions the secondary maxima 
shift a little with increasing excitation energy (as indicated by the dashed 
line in Figure \ref{expad}), but lie within the region 29$^{\circ}$ to 
41$^{\circ}$. We note that the secondary maxima appear to be less strong in 
the case of the 14.1 and 14.6 MeV states.
The observed small change of about 10 degrees of the maxima with excitation 
energy may reflect the changing amplitudes of the two contributing angular 
momentum transfer values. The data thus appear to indicate that the five 
states possess the same value of $L_f$ which may be either 1 or 2.  
From the perspective of the excitation energy of the observed states the 
latter is most likely. However, we note that the angular distributions in
\cite{Bet70} were measured at centre-of-mass energies approximately a factor
of 2 smaller than in the present case, which may result in some variations of
the distributions in  \cite{Bet70} and those measured here.

These data would suggest that the five states at 14.1, 14.6, 16.8, 18.7 and 
21.3 MeV all correspond to a similar angular momentum transfer and thus may 
be structurally linked, whereas the two states at 13.4 and 23.9 MeV may 
possess an alternative character. This observation appears to be inconsistent 
with the assignments made by Milin and von Oertzen.

We have attempted to reproduce the experimental angular
distributions using the DWBA formalism. However, the calculated
distributions were found to be overly sensitive to the details of
the interaction potentials, and moreover required an approximation
to bound state wave-functions for the unbound resonant states.
These features of the calculations have not permitted spin
information to be accurately and unambiguously extracted from the
present data.

\section{Summary}

A measurement of the $^{7}$Li($^9$Be,$^{9}$Be$\alpha$)$^{3}$H at
$E_{beam}$=70 MeV has been performed using an array of charged
particle detector telescopes. The measurement of the $^{13}$C
excitation energy spectrum corresponding to decays into
$\alpha$+$^9$Be produced evidence for excited states between 12
and 28 MeV. Evidence is found for three new broad resonances at
18.7, 21.3 and 23.9 MeV. Given the nature of the reaction process,
$\alpha$ transfer onto the $\alpha$+$n$+$\alpha$ cluster nucleus
$^9$Be, it is possible that some of the states observed in the present
measurement may be linked with the 3$\alpha$+$n$ chain structure
suggested by Milin and von Oertzen \cite{Mil02}. An analysis of
the angular distributions suggests that the group of states at 
excitation energies 14.1, 14.6, 16.8, 18.7 and 21.3 MeV correspond 
to the $\alpha$-transfer to a common orbital in $^{13}$C. The 
measurements suggest that there may be inconsistencies in the 
assignments made in reference \cite{Mil02}, in identifying the 
chain-structure.
Particularly the resonance at 14.6 MeV observed in the present
data is strongly populated, and thus would be expected to feature
strongly in the rotational systematics. Further, the measurements
indicate that the states may not be ordered into the bands
suggested by Table \ref{matkovon}.

Clearly, spin determinations are imperative in order to fully understand the
nature of the states above the $^9$Be+$\alpha$ decay threshold. Due to the
large number of reaction amplitudes, resulting from the presence of non-zero
spin particles, the angular correlations measured here were featureless. It is
possible that reactions such as $^{12}$C($^{9}$Be,$^{13}$C)$^{8}$Be and
$^{16}$O($^{9}$Be,$^{13}$C)$^{12}$C, possessing spin-zero nuclei in the initial
and final state will provide a more promising avenue for resolving this issue.
Alternatively, an analysis of the $^6$Li($^9$Be,$^{13}$C)$^2$H reaction should 
produce angular distributions which are more oscillatory than those observed 
in the present study.

{\bf Acknowledgments}

The authors would like to acknowledge the assistance of ANU personnel in
running the accelerator. This work was carried out under a formal agreement
between the U.K. Engineering and Physical Sciences Research Council and the
Australian National University. PJL, BJG and KLJ would like to acknowledge the
EPSRC for financial support.

\end{document}